\documentclass[11pt]{article}
\usepackage{moriond,epsfig}

\begin{document}
\title{ELECTRONIC EPR-LIKE EXPERIMENTS WITH
SUPERCONDUCTORS}
\author{\underline{R. M\'ELIN}}

\address{Centre de Recherches sur les Tr\`es Basses
Temp\'eratures (CRTBT)\\
CNRS, BP 166 X, 38042 Grenoble Cedex, France}

\maketitle
\abstracts{We discuss a few situations related to
non separable correlations in multiterminal
hybrid structures. We
show that the existence of such
correlations can modify the strength of the
gap of the superconductor. We discuss
linear combinations of non local Cooper pairs.
We discuss a
possible transport experiment intended to
probe non separable correlations.
The models are worked out
at the heuristic level of
effective single site Green's functions.
}

\section{Introduction}

Non locality and non separable correlations
are one of the deepest features of quantum
mechanics. The presence of non separable
correlations in quantum mechanics was identified by
Einstein, Podolsky and Rosen~\cite{EPR} (EPR) in 1935.
One of the possible alternatives to quantum
theory could have been hidden variable theories.
Bell has shown in 1964
that there exists a basic
difference between hidden variable theories
and quantum mechanics~\cite{Bell}.
A. Aspect has shown experimentally
in 1981 that Bell inequalities were violated
with photons~\cite{Aspect}. This constitutes
the ultimate test of quantum mechanics,
which is a non local theory without hidden
variables.

Our goal is to describe the physics
of heterostructures in which a superconductor is
connected to several electrodes, and discuss 
the consequences for electronic
EPR-like experiments. In these systems,
non separable correlations are due to superconducting
pairs correlations~\cite{Loss,Martin}.
The approach is heuristic in the sense that it contains
the correct phenomenology
but does not rely on full mathematical rigor.
We use a single site Green's
function toy-model in which
charge conservation is enforced
``by hand''~\cite{Melin-long}. 
This approach is complementary to
lowest order perturbation theory~\cite{Feinberg}.
We already worked out part of the
rigorous formulation of microscopic transport theory
which will be presented elsewhere~\cite{Melin-Feinberg}.

\section{Implications for thermodynamics}
\label{sec:correlated-pairs}

Let us start to discuss a model in which a superconductor
$x$ is connected to $N$ ferromagnetic electrodes
$\alpha_k$, $k=1, ..., N$~\cite{Melin-short}.
Each of these elements is represented by an effective
single site Green's function~\cite{Cuevas}.
We calculate the Gorkov function
$\left[\hat{G}^{+,-}(\omega)\right]_{1,2}$
of the superconducting site connected to the
ferromagnetic electrodes~\cite{Melin-short}.
The hopping matrix elements can be arbitrarily
large. The gap should satisfy the self-consistent
equation $\Delta = U \int_{-\infty}^{+\infty}
d \omega / (2i\pi) \left[\hat{G}^{+,-}
(\omega)\right]_{1,2}$, where $U$ is the
microscopic attractive interaction.
This leads to the
modified BCS relation
\begin{equation}
\label{eq:gap1}
\Delta = D \exp{ \left[ -  \frac{1}{\rho_N U}
\left( 1 + \pi \rho_N \Gamma_\uparrow \right)
\left( 1 + \pi \rho_N \Gamma_\downarrow \right)
\right]}
,
\end{equation}
where $D$ is the bandwidth, $\rho_N$ is the
normal state density of states and 
$\Gamma_\sigma$ is the spectral linewidth
of spin-$\sigma$ electron:
$\Gamma_\sigma = \sum_{k=1}^N
\Gamma_{k,\sigma}$, where
$\Gamma_{k,\sigma}
= | t_{x,\alpha_k} |^2 \rho_{k,\sigma}$.
The spectral linewidths
measure the coupling between the superconductor
and the ferromagnetic electrodes. With two electrodes
only, one has $\Gamma_\uparrow = 2 \gamma$,
$\Gamma_\downarrow = 0$ in the parallel alignment.
In the antiparallel alignment, one has
$\Gamma_\uparrow = \Gamma_\downarrow
= \gamma$. We deduce from Eq.~\ref{eq:gap1} the
value of the ratio of the gaps:
$$
\frac{\Delta_{AP}}{\Delta_P} = \exp{ \left( -
\frac{\pi^2 \rho_N \gamma^2}
{ U } \right)}
.
$$
The gap is stronger if the electrodes
are in a parallel alignment. This
is because the proximity effect is stronger
in the antiparallel alignment.
This behavior coincides
with a diffusive model
solved recently~\cite{Baladie}.

\section{Linear superpositions of correlated pairs
of electrons}
\label{sec:linear}
\begin{figure}
\centerline{\psfig{figure=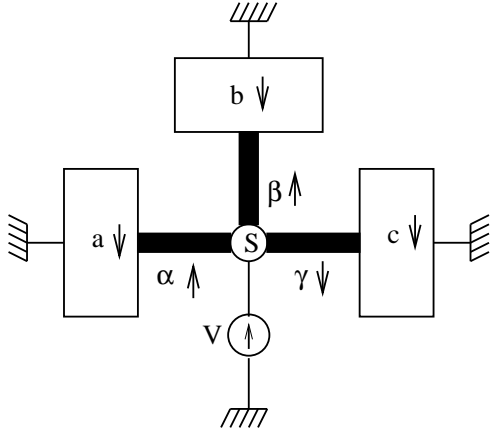,height=2.25in}}
\caption{Representation of a device in which three
ballistic ferromagnetic regions
$\alpha$, $\beta$, $\gamma$
are connected to a superconductor. The additional
electrodes a, b, c are used to
to perform a measurement of the linear superposition.
\label{fig:3ferro}
}
\end{figure}

Now we consider a system in which a superconductor is
connected to three ferromagnetic ballistic regions by
high transparency contacts. The three ferromagnetic
ballistic regions are connected to three ferromagnetic
reservoirs by low transparency contacts (see Fig.~\ref{fig:3ferro}).
We want to show that transport in this system can
be interpreted in terms of projections of a
BCS-like wave function, which is a linear superposition
of correlated pairs~\cite{Melin-long}.

\medskip

\noindent
{\sl Transport formula:}
We assume that a voltage $V$ is applied on the superconductor.
The Andreev current is found to be
the sum of all Cooper
pair transmissions~\cite{Melin-long}:
\begin{equation}
\label{eq:Landauer}
I^A_{n,\sigma} = 4 \pi^2 
\int d \omega \left[ n_F(\omega-eV)
- n_F(\omega) \right] \sum_{m}  
\hat{Q} \left[
u_{n,\sigma,\sigma'}
u_{m,-\sigma,\sigma'}\right]
|G_{x,x,1,2}^A|^2
,
\end{equation}
where
$
u_{n,\sigma,\sigma'} = \pi^2
|t_{x,\alpha_n}|^2 |t_{a_n,\alpha_n}|^2
{\rho_{\alpha_n,\sigma}^2 \rho_{a_n,\sigma}}/
{ \left( 1 + \pi^2 |t_{a_n,\alpha'_n}|^2
\rho_{\alpha_n,\sigma}
\rho_{a_n,\sigma}  \right)^2}
$.
The operator $\hat{Q}$ is used to enforce
charge conservation~\cite{Melin-long}.

\medskip

\noindent
{\sl BCS-like wave function:}
It is tempting to look for a BCS-like wave function
associated to the situation on Fig.~\ref{fig:3ferro}.
For instance in the presence of only two electrodes
$\alpha$ and $\beta$ having a spin orientation
$\sigma_\alpha=\uparrow$ and $\sigma_\beta=\downarrow$,
the superconducting wave function is the product of
two contributions:
(i) Local Cooper pairs (residing in S);
(ii) Non local Cooper pairs (residing in the
ferromagnets $\alpha$ and $\beta$). We drop
out the contribution of local Cooper pairs
and write  the wave function under the form
$|\psi \rangle = c_{\alpha,\uparrow}^+
c_{\beta,\downarrow}^+ | 0 \rangle$.
In the presence of three
ballistic regions $\alpha$, $\beta$, $\gamma$,
the BCS-like wave function takes the form
\begin{equation}
\label{eq:psi}
| \psi \rangle =
c_{\alpha, \uparrow}^+ \left[
\sqrt{ \frac{t_\beta \rho_{\beta,\downarrow}}
{t_\beta \rho_{\beta,\downarrow} +
t_\gamma \rho_{\gamma,\downarrow}}}
c_{\beta,\downarrow}^+
+
\sqrt{ \frac{t_\gamma \rho_{\gamma,\downarrow}}
{t_\beta \rho_{\beta,\downarrow}+ t_\gamma
\rho_{\gamma,\downarrow}}}
c_{\gamma,\downarrow}^+
\right] |0 \rangle
,
\end{equation}
where we assume that $\alpha$, $\beta$ and 
$\gamma$ are three fully spin polarized
ferromagnets with spin orientations
$\sigma_\alpha=\uparrow$,
$\sigma_\beta=\sigma_\gamma=\downarrow$
(see Fig.~\ref{fig:3ferro}).

\medskip

\noindent
{\sl Interpretation of the transport formula:}
In the presence of an arbitrary number of
electrodes, the BCS-like wave function is
a superposition of all possible Cooper pairs:
$$
| \psi \rangle =
{\cal N}^{-1/2}
\sum_{p,q} \sqrt{ t_{\alpha_p}
t_{\alpha_q} \rho_{\alpha_p,\uparrow}
\rho_{\alpha_q,\downarrow}}
c_{\alpha_p,\uparrow}^+
c_{\alpha_q,\downarrow}^+
| 0 \rangle
,
$$
where the normalization coefficient is
${\cal N} = \sum_{p} t_{\alpha_p}
\rho_{\alpha_p,\uparrow} 
\sum_{q} t_{\alpha_q}
\rho_{\alpha_q,\downarrow}
$.
One can define a projection operator associated
to the Cooper pair $(p,q)$:
$
\hat{\cal P}_{p,q}
= c_{\alpha_{p},\uparrow}^+
c_{\alpha_{q},\downarrow}^+
c_{\alpha_{p},\uparrow}
c_{\alpha_{q},\downarrow}
$.
The spin-up current through electrode
$a_p$ is found to be
\begin{equation}
\label{eq:current-proj}
I_p= 4 \pi^2 {\cal N}^2
\sum_{q}
\int d \omega \left[ n_F(\omega-eV)
-n_F(\omega) \right] \gamma_{a_p,\uparrow}
\gamma_{a_q,\downarrow}
\left| \langle \psi | 
\hat{\cal P}_{p,q}
| \psi \rangle \right|^2
|G_{x,x,1,2}^A|^2
,
\end{equation}
where $\gamma_{a_{p (q)},\uparrow} = 
\pi^2 |t_{a_{p (q)}}|^2 \rho_{a_{p (q)},\uparrow
(\downarrow)}$. We interpret the transport
formula (\ref{eq:current-proj}) as follows:
(i) {\sl Before the tunneling event}, the Cooper
pairs are described by the BCS-like wave function.
(ii) {\sl Short after the tunneling event}, there
is one Cooper pair transfered in electrodes
$(a_p,a_q)$. 
(iii) {\sl Long after the tunneling event},
there is a spin-up electron in electrode $p$ and
a spin-down electron in electrode $q$
without phase coherence between the two.

The situation in which S is replaced by
a normal metal is already non trivial. In this
case, the wave function is a linear
superposition of single electron states
while we pointed out here the existence of
a linear superposition of two-electron states.

\section{Experiment proposal}
\label{sec:realistic}
\begin{figure}
\centerline{\psfig{figure=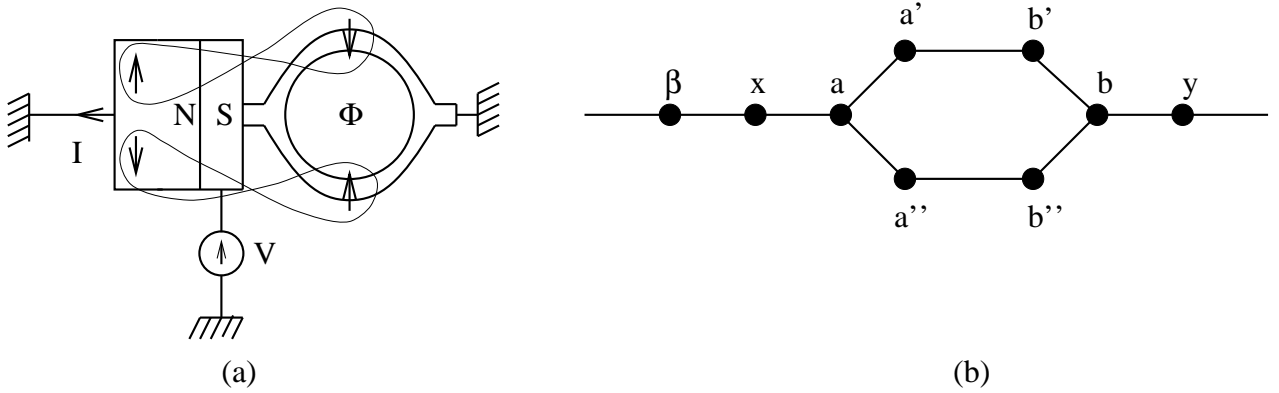,height=2.in}}
\caption{(a) Geometry of the Aharonov-Bohm experiment.
The current $I$ flowing into the left electrode
is modulated by the flux enclosed by the other electron
making the correlated pair. 
(b) The circuit used in the single site version of
Keldysh formalism. Site $x$ is superconducting.
\label{fig:AB}
}
\end{figure}

\noindent
{\sl Geometry:}
Now we discuss a more realistic
experiment that can be made without ferromagnets.
The idea is to connect two electrodes to a
superconductor. One of these electrodes contains an
Aharonov-Bohm loop (see Fig.~\ref{fig:AB}).

\medskip

\noindent
{\sl Physical picture:}
Let us assume
that the Aharonov-Bohm loop is weakly connected
to the rest of the circuit. In the tunnel regime, there
are discrete energy levels $\epsilon_n(\phi)$
on the Aharonov-Bohm loop, which are periodic
functions of $\phi$. One-electron conduction is maximal
when one level is resonant. 
The Aharonov-Bohm loop can be represented by an effective
transmission amplitude $T(\phi)$. The Cooper pair current
is proportional to $|t_{x,\beta}|^2 |T(\phi)|^2$, which
oscillates with $\phi$. As a consequence, the current
through the left electrode on Fig.~\ref{fig:AB}
oscillates as a function of the flux enclosed
in the right electrode. 

\medskip

\noindent
{\sl Single site calculation:}
The non local contribution to the current through the left
electrode on Fig.~\ref{fig:AB} takes the form~\cite{Melin-long}:
\begin{eqnarray}
\nonumber
I^{A,n.l.}_{\beta,\sigma} &=&
\int 
\hat{Q} \left[
\frac{ 4 \pi^2 |t_{x,\beta}|^2 |t_{x,a}|^2
\rho_{\beta,\sigma} \rho_{a,-\sigma}^2
\rho_{b,-\sigma} \left[
(\gamma'_{-\sigma} + \gamma''_{-\sigma})^2
+2 \gamma'_{-\sigma} \gamma''_{-\sigma}
[1 - \cos{(2 \pi \phi / \phi_0)}] \right]}
{ \left[ 1 + (\rho_{a,-\sigma} + \rho_{b,-\sigma})
(\gamma'_{-\sigma} + \gamma''_{-\sigma})
+2 \gamma'_{-\sigma} \gamma''_{-\sigma}
\rho_{a,-\sigma} \rho_{b,-\sigma}
[1 - \cos{(2 \pi \phi / \phi_0)}] \right]^2} \right]\\
&& \times \left[ n_F(\omega-eV) - n_F(\omega) \right]
|G_{x,x,1,2}^A|^2 d \omega
\label{eq:current-nl}
.
\end{eqnarray}
The current (\ref{eq:current-nl})
oscillates with a period $\phi_0=h/e$,
which is in agreement
with the tunnel limit behavior.
The magnitude of the oscillatory component
is of order $|t_{x,\beta}|^2 |t_{x,a}|^2$.
Note that the local current
\begin{equation}
I^A_{\beta,{\rm local}} = 
4 \pi^2 |t_{x,\beta}|^4 \rho_{\beta,\uparrow}
\rho_{\beta,\downarrow} \int
d \omega \left[ n_F(\omega-eV) - n_F(\omega) \right]
|G_{x,x,1,2}^A|^2
,
\end{equation}
also oscillates with $\phi$ because of the renormalization
of the Green's function $G_{x,x,1,2}^A$.
These oscillations in the local contribution
correspond to processes
in which a Cooper pair is extracted from the
superconductor and is transferred into the right
reservoir. The Cooper pair couples to the Aharonov-Bohm
flux,
comes back in the superconductor and is transferred
into the left electrode. This process appears
at order $|t_{x,\beta}|^4 |t_{x,a}|^8$ while the
current due to non separable correlations appears
at order $|t_{x,\beta}|^2 |t_{x,a}|^2$. If at
least one of the interfaces has a low transparency,
it would be possible to retain
mostly the current due to non separable correlations.

\section{Conclusions}
To summarize, we have pointed out a few situations
intended to probe
non separable correlations in electronic systems.
The dependence of the superconducting
gap upon the spin orientation of a ferromagnetic
environment may be tested in experiments.
It would be harder to test the linear superposition
of correlated pairs of electrons. 
The Aharonov-Bohm
effect can be probed in experiments
because phase coherence can propagate over
sufficiently large distances in normal metals.
Finally, we note that an electronic EPR experiment
based on ballistic electrons in semiconductors
has been proposed recently~\cite{Ion}.

\section*{Acknowledgments}
The author wishes to thank fruitful discussions with
P. Degiovanni, D. Feinberg and S. Peysson.

\end{document}